\documentclass[11pt]{article} % use larger type; default would be 10pt

\usepackage[utf8]{inputenc} % set input encoding (not needed with XeLaTeX)
\usepackage{amsmath}
\newcommand{\mynote}[1]{}
\newcommand{\mean}{\mathrm{mean}}

\usepackage{caption}
\usepackage{subcaption}
\usepackage{textcomp}
\usepackage{geometry} % to change the page dimensions
\usepackage{graphicx} % support the \includegraphics command and options
\usepackage{cite} %collapse sequential reference numbers
\usepackage{booktabs} % for much better looking tables
\usepackage{array} % for better arrays (eg matrices) in maths
\usepackage{paralist} % very flexible & customisable lists (eg. enumerate/itemize, etc.)
\usepackage{verbatim} % adds environment for commenting out blocks of text & for better verbatim
\title{Calculation of $\pi$ on the IBM quantum computer and the accuracy of one-qubit operations}
\author{G.A. Bochkin, S.I. Doronin, E.B.Fel'dman, A.I. Zenchuk \\ \small Institute of Problems of Chemical Physics of Russian Academy of Sciences, 142432, \\  \small Moscow Region, Chernogolovka}
\begin{document}

\maketitle

A quantum algorithm for the calculation of $\pi$ is proposed and implemented
on the five-qubit IBM quantum computer with superconducting qubits. We find $\pi=3.157\pm0.017$. The error is due to the noise of quantum one-qubit operations and measurements. The results can be used for estimating the errors of the quantum computer and suggest that the errors are purely random.

\section{Introduction}\label{sec:intro}
Many problems were considered for the IBM quantum computer with superconducting qubits \cite{PhysRevA.76.042319,chow}, but high levels of noise preclude solving most of them% \cite{}
. The problem of decreasing the noise and increasing the computation accuracy remains very important. Various schemes for solving the noise problem were considered \cite{PhysRevLett.119.180509,kandala2,corcoles}. There were also some proposals for the tasks suitable for existing quantum computers \cite{zhukov,doronin2019solving,kandala,PhysRevX.7.021050}.  

In this paper, we do not try to best classical computers with a quantum one. Instead, we focus on a classical problem that has long been used to highlight the capabilities of classical computers and the power of applied mathematics in the era before computers. The problem consists in computing the number $\pi$.

Arguably, Archimedes' results for $\pi$ from 23 centuries ago have been sufficiently accurate for many applications. Nevertheless, a truly gigantic number of digits has been computed since then. We do not attempt to beat the existing records and focus on getting the best result for $\pi$ we can extract with the IBM quantum computer.

Our algorithm uses only one-qubit operations. We apply our results to estimate the accuracy of the quantum computer. The precise error rate of an early computer might not be of great interest, but  the nature of its errors is. Are the dominant errors random or systematic? Our findings are consistent with purely random errors.

This article is organized as follows. In Sec.~\ref{sec:idea}, the idea of the computational method is presented. The algorithm for calculatiing $\pi$ on a quantum computer is given in Sec.~\ref{sec:algo}. We briefly summarize our results in the concluding Sec.~\ref{sec:concl}.

\section{The idea}\label{sec:idea}
Let us consider a qubit in the initial state $\left|0\right>$. After ``rotating'' it by $\varphi$ radians (on the Bloch sphere) around the $y$-axis, its state will be $\exp(-iI_y\varphi)\left|0\right>$.  ($\varphi$ is thus the dimensionless time). If it is subsequently measured in the basis  $\{\left|0\right>,\,\left|1\right>\}$, then the outcome will be $\left |0 \right>$ with the probability  $\cos^2(\varphi/2) = (1 + \cos \varphi)/2$  and $\left |1 \right>$ with the probability 
\begin{equation}p(\varphi)=\sin^2(\varphi/2)=(1 - \cos \varphi)/2\label{theorpval}. 
\end{equation}

The difference between any two successive roots of the equation $p(\varphi)=1/2$, for example, $\pi/2$ and $3\pi/2$, is exactly $\pi$. 
Hence, $\pi$ can be determined from the knowledge of $p(\varphi)$.
Of course, we cannot measure $p(\varphi)$ exactly. On a real quantum device, we cannot even guarantee that the rotation angle is exactly as desired. We expect, however, that the real rotation angle $\varphi = ct$, where $t$ is a value which can be precisely controlled (such as the duration of a control pulse) and $c$ is a constant. 
Thus, we can estimate $p(ct)$ for a given $t=\varphi/c$ as the fraction $f(t)$ of the measurements that give the result $\left|1\right>$. To proceed further with calculating $\pi$, we need to estimate the value of $c$ as well. 

\subsection*{Estimation of the constant $c$}
Below, we will use the notation $t_1 = \pi/2c$ and $t_2=3 \pi/2c$. These are the first two nonnegative roots of the equation $p(ct)=1/2$. 
Since \begin{equation}\int\limits_{t_1}^{t_2}\left(p (ct)-\frac 12\right)\,dt =%\int\limits_{\pi/2c}^{2\pi/2c}\left(p (t)-\frac 12\right)\,dt= \int\limits_{\pi/2c}^{2\pi/2c}\left(\frac12 (1-\cos (ct))-\frac 12\right)\,dt=\frac 1c\int\limits_{\pi/2}^{{3\pi}/2}-\frac12 \cos (x)\,dx= 
\frac 1c\label{cmethod},
\end{equation}
(see Eq. \eqref{theorpval}), the area under the experimental curve $\left(f(t)-\frac 12\right)$ on the interval $[\hat t_1,\hat t_2]$ (where the hats denote an estimate of the respecitve variable) provides a way to estimate $c$.

Due to the experimental imperfections,  $\left|0\right>$ and  $\left|1\right>$ cannot be obtained with certainty, no matter what operations are performed.
For the sake of estimating errors, we approximate the experimental probability of obtaining the result  $\left|1\right>$ as
\begin{equation}
P(t) = \alpha \frac{1-\cos (ct +\phi_0)}2 +\beta \label{pfudge},
\end{equation}
with three empirical constants $\alpha$, $\beta$, $\phi_0$ in addition to $c$. That is, $P(t)$ is derived from $p(\varphi)$ by arbitrary linear transforms of both its domain (with parameters $c$ and $\phi_0$) and the image (with parameters $\alpha$ ,$\beta$).
These empirical parameters allow us to account for %\begin{itemize}
the initial state not being  exactly $\left|0\right>$ (it could even be mixed) and %; see then next item).
measurement errors.  In the ideal case $\alpha = 1$, $\beta = \phi_0 = 0$.  In any case $\alpha \ge 0$, $0<\beta<1-\alpha$, since the probability must be between 0 and 1.

\section{The algorithm}\label{sec:algo}
We assume that the approximate period of $P(t)$ is already known, and time units are chosen so that it is approximately 6.
The fraction $f(t)$ will often be called simply ``fraction''. Let $T$ be the set of all values of $t$ for which the experiments were performed.%, $t_{min} = \min T$,  $t_{max} = \max T$.
The algorithm consists of the following steps:
\begin{enumerate}
\item Roughly estimate $\alpha$ and $\beta$  (the measurement imperfections): find $\hat\alpha$, $\hat\beta$ from the system of equations
\begin{equation}\left\{\begin{aligned}
\min\limits_t f(t) &= \hat\beta \\
\max\limits_t f(t) &= \hat\alpha+\hat\beta
\end{aligned}\right.\label{simplealphaandbeta}\end{equation}

\item\label{it:normalize} Normalize $f(t)$ so that its minimum is 0 and maximum is 1:
$$f_1(t) = \frac{f(t)-\hat\beta}{\hat\alpha}$$
\item Define $\tilde f_1(t)$ as the extension of $f_1$ for values of $t$ between $\min T$ and $\max T$ by linear interpolation between neighboring values of $f_1(t)$.
\item Roughly estimate $t_1$ and $t_2$: find two time instants $t$ such that $\tilde f(t)= 0.5$ using root search methods, $\hat t_1$ starting from the point $t=1.5 $ and $\hat t_2$ starting from $t=4.5$.
\item\label{refinealphaandbeta} Refine the estimate for $\alpha$ and $\beta$: solve the system \begin{equation}\left\{\begin{aligned}
\mean &\left\{f_1(t)\middle|\,|t-\hat t_{minval}|<\delta\right\}  =& \hat\beta & \\
\mean &\left\{f_1(t)\middle|\,|t-\hat t_{maxval}|<\delta\right\}  =& \hat\alpha+\hat\beta &
\end{aligned}\right.\label{refinedalphaandbeta}\end{equation}
where $\mean\, S$ is the arithmetic mean of the set $S$, $\hat t_{maxval} = \frac{\hat t_1+\hat t_2}2$ is the estimated maximum point of $P(t)$ calculated from $t_1$ and $t_2$, similarly for $\hat t_{minval}$; $\delta$ is chosen so that $1-\cos(c\delta) \ll {\sqrt{P(t)(1-P(t))}}$. (The right-hand side is the standard deviation of a Bernoulli distribution.)
\item Repeat step~\ref{it:normalize}
 with new $\hat\alpha$ and $\hat\beta$. 
\item Refine the estimate for $t_1$ and $t_2$ as follows. For $i = 1, 2$:\begin{enumerate}
\item take the experimental times $t$ such that \mbox{$|t-t_i|\le 0.5$} and corresponding $f_1(t)$\item fit  a linear function $\gamma t+k$ to these $f_1(t)$ by the least-squares method
\item the new $\hat t_i$ is a solution to the equation $\gamma\hat t_i+k=0.5$ \end{enumerate}
\item Find the integral $I=\int\limits_{t_1}^{t_2}\left(f (t)-\frac 12\right)\,dt$ using the trapezoidal rule.
\item Estimate $\pi$ as $\frac{t_2-t_1}I$.
\end{enumerate}

\begin{figure} 
\begin{subfigure}{0.4 \textwidth}\includegraphics{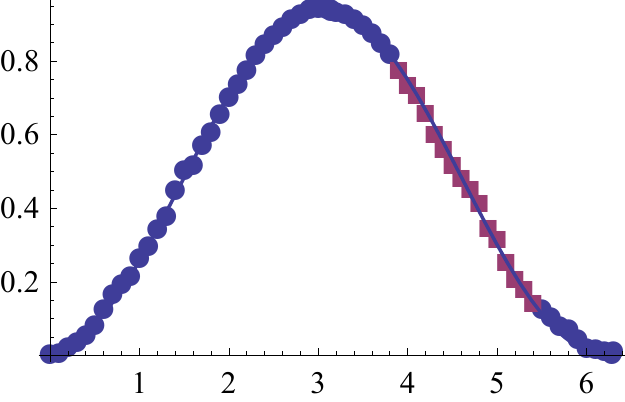}\caption{Qubit \#1}\end{subfigure}
\begin{subfigure}{0.4 \textwidth} \includegraphics{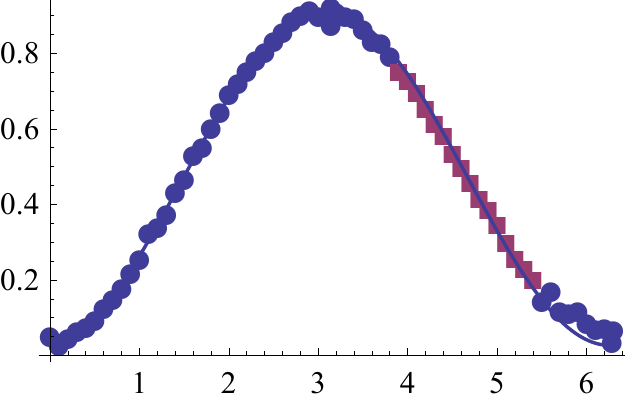}\caption{Qubit \#2}\end{subfigure}
\begin{subfigure}{0.4 \textwidth}\includegraphics{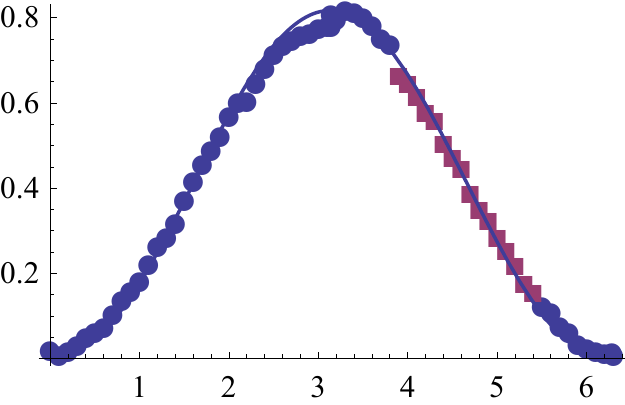} \caption{Qubit \#3}\label{sfig:q3}\end{subfigure}
\begin{subfigure}{0.4 \textwidth}\includegraphics{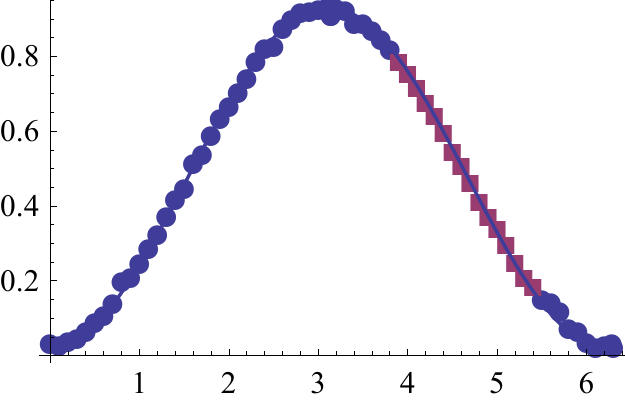} \caption{Qubit \#4}\label{sfig:q4}\end{subfigure}
\begin{subfigure}{0.4 \textwidth} \includegraphics{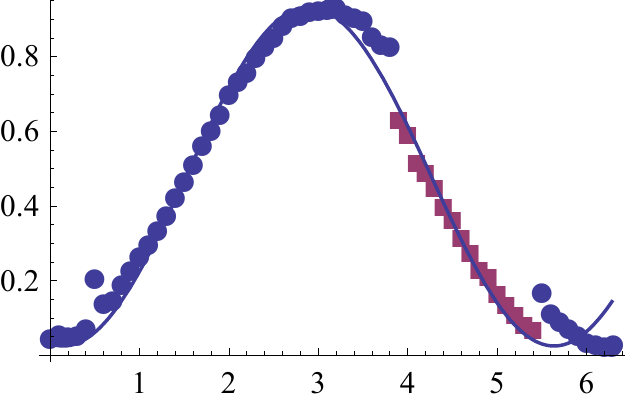} \caption{Qubit \#5}\label{sfig:q5}\end{subfigure}
\caption{Experimental fractions of $\left|1\right>$ measurement outcomes (points) and curves obtained by fitting parameters in Eq.~\eqref{pfudge} (solid line). Qubits~\#1~(a), \#2~(b), \#3~(c), \#4~(d) and \#5 (e). Horizontal axis is the rotation angle $t$}
\label{fig:q1to5}
\end{figure}

We made 8192 measurements on each of the five qubits for each time instant from 0 to 6.3 with the step of 0.1. The value $\delta =0.1$ (see step \ref{refinealphaandbeta}) was chosen.  
Each value of $t$ required a separate job on the quantum platform. 
The results on qubit \#5 (Fig.~\ref{sfig:q5}) contain a large jump near $t=4$, apparently due to a calibration difference between our runs. In the graph for qubit \#3 (Fig.~\ref{sfig:q3}), a step near the time instant $t=\pi$ can also be observed. The results from those qubits were discarded. The results on the three remaining qubits are well described by Eq.~\eqref{pfudge} in the region of interest ($1<t<5$). The dependence of the fractions on $t$ is plotted in Fig.~\ref{fig:q1to5}. On the three plots corresponding to those qubits, the points are, for the most part, close to the theoretical curve.
\subsection*{Checking the algorithm's accuracy}
In order to check the accuracy of the algorithm, we estimated parameters $\alpha$ and $\beta$ for the three qubits used in the experiment from experimental data; generated synthetic data with the probabilities according to Eq.~\eqref{pfudge} for time instants $t$ as in the experiment (see above); ran the algorithm 50 times for each of three (estimated) pairs $(\alpha, \beta)$, for a total of 150 times; and calculated the standard deviation of the result. 

This also allows to measure the impact of random errors on any intermediate result produced by the algorithm, in the same way. We found that the standrard deviation of $(t_2-t_1)$ is 0.009 and the standard deviation of $I$ is 0.006.

%It must be noted that the errors in estimating $(t_2-t_1)$ and $c$ cannot be considered independent. Therefore, the estimation error for $\pi$ cannot be calculated from estimation errors for $(t_2-t_1)$ and $c$. Instead, it was computed numerically using the same general idea.
The comments on the accuracy of the algorithm below and our conclusion that random errors exceed systematic ones are based on this calculation.

After performing the calculation with the above algorithm and averaging the results from different qubits, we obtain the result $\pi \approx 3.157\pm0.017$ (2 standard deviations calculated as described above).

\subsection*{Discussion}
Using more complicated integration rules such as Simpson's rule is pointless in our case, as it won't be more accurate. In the case $c=1$, calculating the theoretical value of integral $I$ by the trapezoidal rule with time step of 0.1 gives the value of 0.99917 instead of the exact value $I=1$. Thus, the error due to the trapezoidal rule is small compared to the error due to limited number of measurements in our case (standard deviation is about 0.006, with the magnitude of measurement errors we observed; it would be 0.0023 in the ideal case) or the estimation error of $(t_2-t_1)$.

A possible way to improve the accuracy of the results would be to measure several periods of $p(t)$ instead of half a period, assuming that decoherence is small enough.
\section{Conclusions}\label{sec:concl}
We suggested a simple quantum algorithm for calculating $\pi$ on a quantum computer and implemented it on the IBM quantum platform. 
We provide an estimate of the magnitude of random errors and show that our results are consistent with the hypothesis of errors being purely random. The computed value of $\pi=3.157$ is within its error bars $\pm 0.017$ from the correct value. This accuracy is still behind the celebrated result $3\frac{10}{71}<\pi<3\frac{1}{7}$ by Archimedes, but it seems that the achievements of classical antiquity are within reach of modern quantum information science.
\subsection*{Acknowledgements}
This work was performed as a part of a state task (State Registration No. 0089-2019-0002).
This work is partially supported by the Russian Foundation for Basic Research (grants \textnumero~19-32-80004 and \textnumero~20-03-00147).
The authors are grateful to D. E. Feldman for useful discussion.
\bibliographystyle{unsrt}
\bibliography{bibl}
\end{document}